# Solitary wave structure of transitional flow in the wake of a sphere

Lin Niu [1], Hua-Shu Dou [1,*], Changquan Zhou[1,2], Wenqian Xu[1,3]

[1]Faculty of Mechanical Engineering, Zhejiang Sci-Tech University, Hangzhou, Zhejiang, 310018, China
[2]School of Mechanical Engineering, Zhejiang University of Water Resources and Electric Power, Hangzhou, Zhejiang, 310018, China
[3]School of Mechanical Engineering, Hangzhou Dianzi University, Hangzhou, Zhejiang 310018, China
*Email: huashudou@zstu.edu.cn

**Abstract** The soliton-like coherent structure (SCS), which has been verified to exist in both transitional and turbulent boundary layers[1-4], still poses a challenge in the understanding of its formation and behavior. In our previous study (Niu et al.[5]), the SCS was also found to exist in the transitional wake flow behind a sphere. In present study, the formation and evolution of the SCS is further investigated at four Reynolds numbers by numerical simulation. The results show that at the early stage of the turbulence transition, the SCS appears as a form of wave packet during the Tollmien-Schlichting (T-S) wave stage. With the increase of the Reynolds number, the SCS reaches its maximum amplitude downstream where the velocity discontinuity occurs. This position is located after the breakdown of the T-S wave and the three-dimensional structure is formed. Then, the SCS conserves its shape and amplitude over a long distance downstream. The relationships among the SCS, the spikes, the vortex structures, and the high-shear layers are further analyzed. It is found that the SCS in the wake flow has similarities to the phenomena observed in boundary layer flows during the turbulent transition. The vortex structures and high-shear layers mostly wrap around the border of the SCS. The vortex structure is considered to be as a consequence of the development of the SCS rather than its cause.

## 1. Introduction

In the 19th century, the British scientist Russell observed a fast-moving boat on a canal coming to a sudden stop, while the mass of water it was propelling formed a large, well-defined packet of water that continued to move at a constant speed. Russell followed this packet of water and observed it moving steadily across the surface of the water until it dissipated after several kilometers. He named this unique water wave phenomenon "solitary wave"[6]. A solitary wave is a special type of fluctuating phenomenon that retains its shape, amplitude and speed as it propagates. This phenomenon has shown important applications in many fields, such as fluid dynamics, optics, plasma physics and biology. In the field of fluid dynamics, researchers have gradually explored the elaborate details of solitary wave behavior in the context of turbulent transitions. They first identified the phenomenon of spikes when the boundary layer flow was undergoing turbulent transition, and then revealed the solitary wave properties of these spikes[1, 2, 4].

When the boundary layer flow on a flat plate undergoes turbulent transition under low disturbance, it has been discovered that after the instability of Tollmien-Schlichting (T-S) waves, there exist three types of turbulent transitions, which are respectively K-type, O-type, and N-type transitions[4, 7, 8]. As early as 1962, Klebanoff et al.[9] found that flash spikes in the K regime are commonly attributed to laminar boundary layer breakdown and flow randomization. Several researchers then provided evidence for the solitary wave properties of the spikes and the relationship linking the spikes to the coherent structures. Orszag and Patera[10] were the first to describe the coherent structure of turbulence as nonlinear wave packets made up of two-dimensional and three-dimensional wave packets interacting resonantly. Cohen et al.[11] focused on exploring the transformation that low-amplitude wave packets, originating from air pulses of short-duration under controlled circumstances, underwent to turn into a turbulent spot within a boundary layer.

Subsequently, the existence of the solitary wave structure has been verified in both transitional and turbulent boundary layers[1-4]. Kachanov's team and Lee's team have conducted detailed studies of the solitary wave in the turbulent transition[1, 2, 4]. Based on experimental and theoretical analysis, Kachanov[12] proposed a theoretical concept for the early phase of the K-regime, the wave-resonance concept of breakdown. Later on, Kachanov[1] provided a comprehensive summary of the solitary wave features of spikes. He pointed out that there are spikes detected in the boundary layer flow, and that these spikes are coherent structures characterized by a deterministic and periodic nature, i.e. coherent structure solitons (CS-solitons). These spikes are represented as ring-like vortices[13]. In different from Kachanov's CS-soliton (ring-like vortices), Lee et al.[2, 3, 14] further studied the behavior of T-S waves and the flow structures in the boundary layer, and discovered that solitons/like coherent structures (SCSs) are actually three-dimensional waves. The evolution of this structure stems from the interaction of the two oblique waves and it significantly influences on transition to turbulence. As the SCS evolves, they can induce strong shear layers that form at the SCS borders and can develop into secondary closed vortices. In terms of the signatures of SCS structure formation, Kachanov's CS-solitons are the ring-like vortices and Lee's SCS is the three-dimensional wave[1, 2]. Using hot-wire measurements, Kachanov's team observed periodic spikes, leading to the formation of SCS structures. Meanwhile, Lee's team repeated Hama and Nutant' experiments[15], using the flow visualization method with hydrogen bubble technique in boundary layer flow under natural transition conditions. They found that the lines of the hydrogen bubbles were parallel until a kink structure was formed. However, when the kink structure appeared, the bubbles first maintained their parallel shape and then developed into a three-dimensional (3D) wave structure that emerged in front of the Λ vortex. They suggest that this 3D wave structure constitutes the natural origin of the Λ vortex and call it the soliton coherent structure (SCS) due to the slow evolution of this wave and the generation of 3D localized flows.

In recent years, Lee and Jiang[4] have discussed the important role of SCSs during the boundary layer transition and their interrelationships with other vortex structures. Feng et al.[16, 17] demonstrated that a travelling wave has the ability to trigger solitary waves while the wave energy is oscillating in the boundary layer. Jiang et al.[18] further analyzed the evolution of SCSs in the K, N and O regimes, and proposed that the low-speed streaks (LSSs) may be composed of several SCSs. In addition, Jiang et al.[19] pointed out that the low-speed streak changes and appears as a 3D wave accompanied by ejection and sweeping behaviors in a turbulent boundary layer. They indicated that the amplification of the 3D wave plays a principal role in driving the development of LSSs. Jiang et al.[20] had also confirmed the existence and evolution of SCS in increasingly turbulent stratified shear layers.

Thus far, studies of the SCS in the turbulent transition have mainly focused on the wall-bounded flows including the channel flow and boundary layer flow[1-4, 16, 17, 19, 20]. In contrast, there are relatively few studies of SCS in free shear flow. Dou[21] showed that the mechanism of turbulence generation and the structure of turbulence in free shear flows are the same as those in wall-bounded flows. They are uniquely determined by the singularity of the Navier-Stokes equation. As such, it is reckoned that the soliton-like structure found in wall-bounded flows in Lee et al.[2, 3, 14] should also exist in free shear flows such as the wake flow. The SCS is very important in turbulent generation. It is necessary to study the mechanism and the structure of this type of soliton-like structure in depth to understand the mechanism of turbulence generation.

Niu et al.[5] discovered the SCS in their study of the turbulence generation mechanism in the wake flow of a sphere at Re=350. In the present paper, the characteristics and evolution of the SCS in the wake of a sphere are further explored through Large Eddy Simulation (LES) for various Reynolds numbers. The function of the SCS within the turbulent transition is investigated as the Reynolds number varies. Further investigations are made on the relationships among the SCS, spikes, high-shear layers, and vortex structures. We apply the $\widetilde{\Omega_R}$ method[22] to acquire the vortex structures of the wake flow, which can effectively capture vortices of different intensities from weak to strong. The explicit expression of $\widetilde{\Omega_R}$[22] is given as

$$\widetilde{\Omega_R} = \frac{\beta^2}{\beta^2 + \alpha^2 + \lambda_{cr}^2 + \frac{1}{2}\lambda_{cr}^2 + \epsilon}, \tag{1}$$

Here, the specific definition of the parameters in Eq. (1) can be found in Liu and Liu[22].

## 2. Numerical simulations

The computational domain taken into account in this paper is illustrated in Fig. 1. A detailed description of the size of the computational domain, the governing equations and numerical method used in this study can be retrieved from Niu et al.[5].

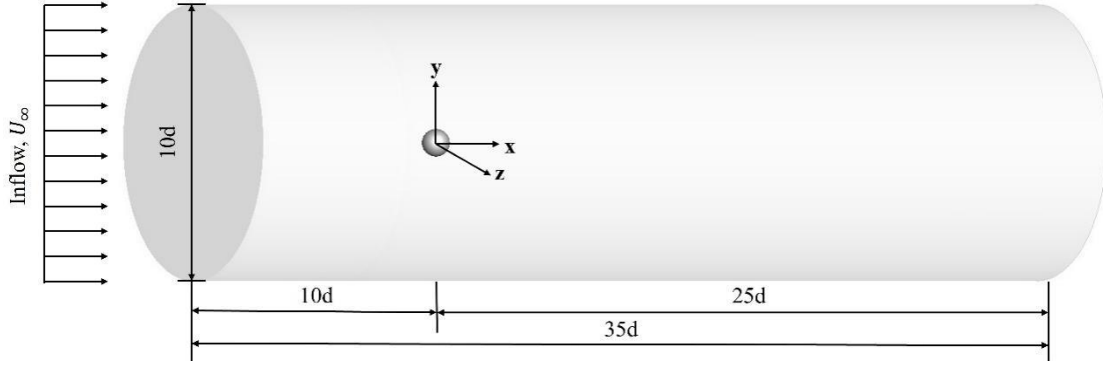

**FIG. 1.** Schematic of the geometry of the computational domain.

For accurately capturing the flow details around the surface of the sphere, we have implemented O-splitting and refinement of the mesh around the sphere to make sure that the y+ value is less than 1 for high resolution. This well-designed mesh layout allows us to accurately represent the subtle flow phenomena around the sphere. In order to further ensure the accuracy of the flow characteristics close to the surface of the sphere, we have progressively refined the mesh around the sphere. Three sets of mesh configurations with different densities have been created, all of which kept the y+ value around the wall below 1 to meet the high-resolution requirement. The numerical results obtained from the three sets of mesh are presented in Table I. The drag coefficients calculated at different mesh densities agree well with the data in the existing literature, thus confirming the accuracy of our mesh design. Taking into account the computational accuracy and efficiency, the mesh with a medium size is used.

**TABLE I.** Impacts of the Reynolds number on the drag coefficient in the flow around a sphere.

| Re | 250 | 270 | 300 | 350 | 500 | 700 | 1000 |
|---|---|---|---|---|---|---|---|
| Coarse mesh | | | | 0.64 | | | 0.52 |
| Medium mesh (Present) | 0.71 | 0.69 | 0.67 | 0.63 | 0.57 | 0.52 | 0.47 |
| Refine mesh | | | | 0.63 | | | 0.48 |
| Wu and Faeth[23] | | | | | | | 0.5 |
| Magnaudet et al.[24] | | | | | | | |
| Mittal[25] | 0.68 | | | 0.62 | | | |
| Johnson and Patel[26] | | | 0.656 | | | | |
| Lee[27] | | | | 0.61 | 0.54 | | |
| Bagchi and Balachandar[28] | 0.7 | | | 0.62 | 0.56 | | |
| Luo et al.[29] | 0.757 | | 0.654 | | | | |
| Mimeau et al.[30] | | | 0.673 | | | | 0.485 |
| Barati et al.[31] | | | 0.65 | 0.622 | 0.55 | 0.5 | 0.46 |
| Tiwari et al.[32] | 0.71 | | 0.67 | 0.63 | 0.58 | | 0.48 |
| Rodrigues et al.[33] | | | | | | | 0.466 |

To check the precision and reliability of the numerical method, a comparative study regarding the drag coefficient and the pressure coefficient has been carried out. The drag coefficient gets treated as $C_d = F_x / \frac{1}{2}\rho U_\infty^2 \pi D^2/4$. Specifically, in this formula, $F_x$ represents the force in the flow direction x, D denotes the diameter of the sphere, $U_\infty$ represents the velocity of the incoming flow, and $\rho$ symbolizes the density. Moreover, the drag coefficient in this study is based on the time-averaged value calculated once the flow reaches a steady state. Table I displays a detailed comparison of the drag coefficients in this study and those sourced from the literature with respect to different Reynolds numbers. The outcomes demonstrate that there exists a remarkable consistency between the drag coefficients obtained in this study and those in the literature, thus verifying the reliability of the numerical method. The pressure coefficient is denoted by $C_p = (p - p_\infty)/\frac{1}{2}\rho U_\infty^2$. Here $p_\infty$ represents the reference pressure at the inlet. The profile of the pressure coefficient $C_p$ along the surface of sphere is illustrated in Fig. 2. Evidently, the current results match those in the references rather well.

Furthermore, the non-dimensional root-mean-square of the streamwise velocity ($u_{RMS}$) as well as the non-dimensional averaged streamwise velocity ($u_{mean}$) at the center line is clearly shown in Fig. 3. This result not only intuitively reflects the flow characteristics, but also shows a high degree of consistency with the findings in the existing literature. Taking into account the distributions of the drag coefficient, the pressure coefficient, the non-dimensional root-mean-square of the streamwise velocity and the non-dimensional averaged streamwise velocity at the central line, the current results agree well with the literature data, thus confirming the accuracy and reliability of our research methodology.

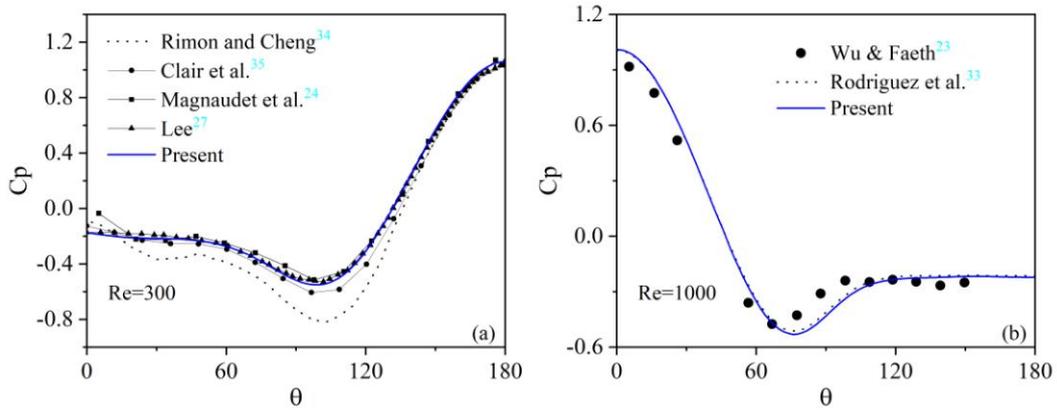

**FIG.2.** Pressure coefficient Cp around the sphere: (a) Re=300, (b) Re=1000. Comparison with results in literature: Rimon and Cheng[34], Clair et al.[35], Magnaudet et al.[24], Lee[27], Wu and Faeth[23], Rodriguez et al.[33].

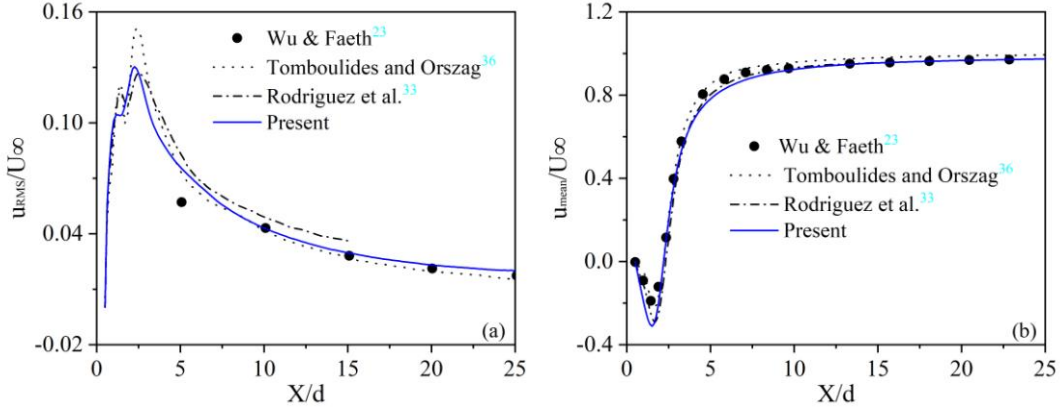

**FIG.3.** (a) Non-dimensional RMS streamwise velocity at the central line of the wake, for Re=1000, (b) non-dimensional averaged streamwise velocity at the central line of the wake, for Re=1000. Comparison with results in literature: Wu and Faeth[23], Tomboulides and Orszag[36], Rodriguez et al.[33].

## 3. Simulation results of flow in the wake

### 3.1 "Kink" structure and 3D wave packet generation in the wake

The vortex structures within the wake identified using the $\widetilde{\Omega_R}$ method at Re=250 are shown in Fig. 4. These structures appear as two parallel vortices in the flow direction. The downwash force generated by these vortices causes the lines passing the vortex cores not to be along the diameter of the sphere, but rather to be biased to one side of the sphere. Comparatively, the vortex structures show a remarkable likeness to that observed by Magarvey and Bishop[37] in liquid-liquid environments, where the streamwise vortices are adjacent to the recirculation zone. Conversely, the vortex structures demonstrated by Tiwari et al.[32] using the Q criterion are not immediately adjacent to the recirculation zone. This suggests that the $\widetilde{\Omega_R}$ method is more effective in capturing weaker vortex structures at low Reynolds number conditions. Moreover, it is evident that the amplitude of the streamwise velocity disturbances within the wake stays quite small, with the maximum value not exceeding 0.001% of the incoming velocity.

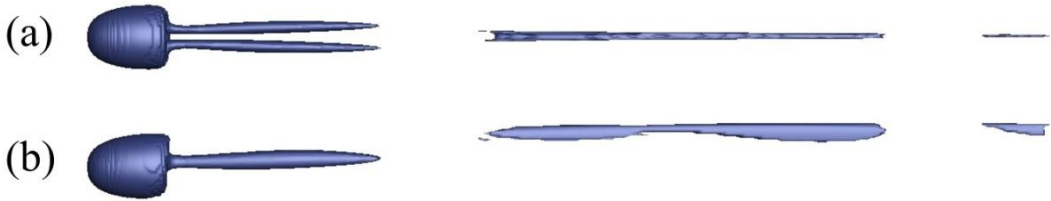

**FIG. 4.** Vortex structures identified by $\widetilde{\Omega_R}$ at Re=250, t=152s: (a) top view, (b) side view.

As the Reynolds number increases, the vortex structures within the wake of a sphere gradually begin to evolve. Fig. 5 shows that the formation of a distinct "kink" structure between two parallel vortices at Re=270 occurs. Moreover, the black circles mark the key stages in the formation of the kink structure. Between the two parallel vortices, wave packets composed of positive streamwise velocity disturbance (red) and negative streamwise velocity disturbance (blue) start to propagate from the recirculation zone and appear as an alternating pattern. At x/d=6, a distinct kink structure is formed above the two streamwise vortices, which then gradually decays along the streamwise direction. At this Reynolds number, the shear layer maintains attached to the sphere all the time.

As shown in Fig. 6, the strength of the vortex in the upper region is relatively weak, while the strength of the vortex in the lower region is slightly stronger. It can be seen that the cores of the two recirculation vortices are always firmly attached to the recirculation zone without separation. Along the streamwise direction, a kink is first observed downstream of the weaker upper recirculation zone. In contrast, the stronger lower recirculation zone does not show any kinks close to the sphere, but rather an expanding vortex in the more distant downstream region. The existence of this difference implies that the shear layer in the upper region is stronger than that in the lower region, which coincides with the layout of vorticity strengths presented in Fig. 6.

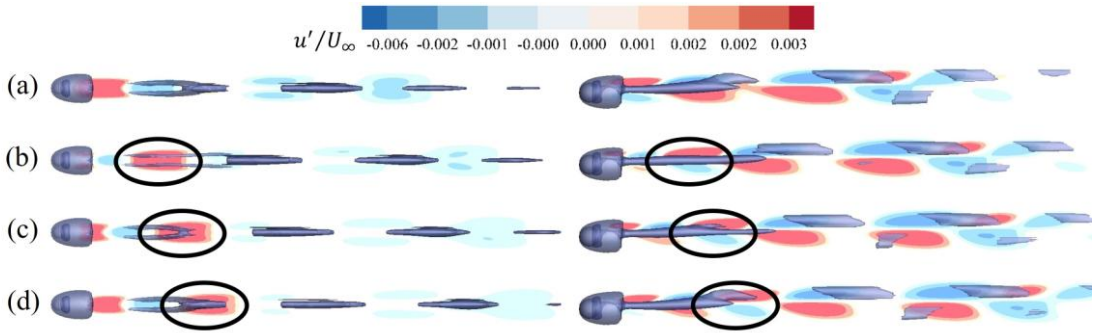

**FIG. 5.** Generation of kink structures in the wake at Re=270, vortex structures identified by $\widetilde{\Omega_R}$, left: top view, contours of $u'/U_\infty$ on the z/d=0.4 plane, right: side view, contours of $u'/U_\infty$ on the y/d=0 plane. (a) t=55s, (b) t=65s, (c) t=69s, (d) t=73s.

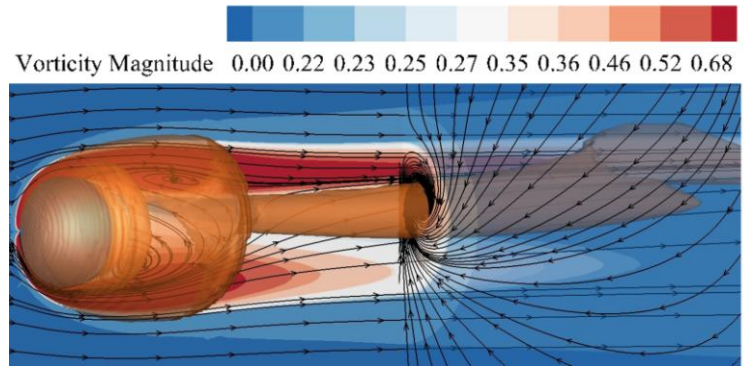

**FIG. 6.** Streamlines in the wake at Re=270, vortex structures identified by $\widetilde{\Omega_R}$ (Orange), contours

of vorticity magnitude on the y/d=0 plane and the x/d=3.937 plane, t=73s.

During the kink formation, the outer side of the positive velocity disturbance region (red) is adjacent to the mainstream region, while its inner side is adjacent to the negative velocity disturbance region (blue). As a result, significant velocity gradients are generated at the boundaries of the travelling positive velocity disturbance region, and shear layers are formed at their junctions, as shown in Fig. 6. As the time increase, these shear effects accumulate in the downstream of the wake, eventually contributing to the formation of the kink structure. In particular, it can be seen that the shear layer attached to the sphere' surface does not separate during this formation stage. This observation suggests that the kink structure does not stem from the shear layer shedding but rather results from the instability of the recirculation zone. This instability leads to velocity disturbances, and these velocity disturbances cause local velocity distortions, which ultimately contributes to the formation of the kink structure. Therefore, the kink structure is a sign of the wave packets of the velocity disturbance. In this evolution, the initial velocity disturbance in the wake originates in the recirculation zone located at the rear of the sphere. Meanwhile, under the influence of this disturbance, the flow itself remains a laminar.

The phenomenon of the kink structure has already been found in a similar Reynolds number range in previous studies[38, 39]. Gumowskiet al.[38] observed the vertical velocity fluctuations from the Reynolds numbers 262 to 291 through visualization experiments. In particular, they recorded harmonic fluctuations in the vertical velocity fluctuation at Re=266 and hypothesized that such fluctuations are associated with the "kink" phenomenon. Thompson et al.[39] further suggested that the kink phenomenon serves as an early sign before the periodic appearance of the hairpin vortices. Furthermore, the kink structures have been found in studies on turbulence onset within the boundary layer flow. Lee[2] repeated Hama and Nutant's 1963 experiments[15] under natural transition conditions. The results of his experiments demonstrated that the kink is a 3D wave structure, which interacts with the boundary layer to form a high-shear layer, and this ultimately results in the generation of a vortex.

As it comes to the kink generation stage, the streamwise velocity fluctuation mainly shows up as a T-S wave with an amplitude that is rather small, around 0.25% of the incoming velocity. In contrast, the velocity fluctuations in the other two directions in the wake are even much smaller in amplitude, with amplitudes of about 0.07% and 0.06% of the incoming velocity. Cohen et al.[11] noticed in boundary layer experiments that during the linear stage of the wave packet's evolution, the amplitude amounts to around 0.3% of the incoming velocity, and then it grows significantly to reach 5% in the subharmonic stage. These findings suggest that the velocity fluctuations within the wake at Re=270 already exhibit three-dimensional properties. Although these three-dimensional properties are relatively weak, they have some similarities to those properties of the boundary layer flow. The wave packets with 3D velocity disturbances represent the initial phase of the SCS and indicate the onset of a more complex flow structure.

## 3.2 SCS generation

The vortex structures at Re=280 are displayed in Fig.7. It appears that the hairpin vortices begin to shed at regular intervals when the Reynolds number reaches 280. Although these vortices are not yet fully matured, they exhibit the basic structure of vortex heads and legs as well as periodic shedding. Therefore, the critical Reynolds number for the formation of hairpin vortices in this study is approximately 280. This conclusion is consistent with the outcomes of several studies. Sakamoto and Haniu[40] determined the critical Reynolds number for hairpin vortex formation to be $R_{e_c} = 300$ using flow visualization and spectral analysis methods. Tombouliders et al.[41] obtained a critical Reynolds number range between 250 and 285 using the large eddy simulations. Wu and Faeth[23] measured experimentally a critical Reynolds number $R_{e_c}$ of 280. Ormieres and Provansal[42] achieved $R_{e_c} = 280$ through a detailed experimental study over a narrow Reynolds number range (100-360).

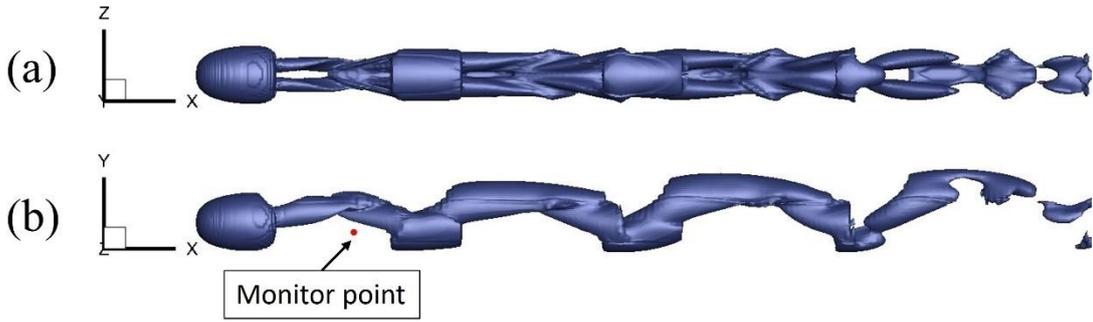

**FIG. 7.** Vortex structures identified by $\widetilde{\Omega}_R$ at Re=280, t=254s: (a) top view, (b) side view.

Fig. 8 clearly demonstrates the regular 3D wave packet morphology presented by the velocity disturbances in the wake at Re=280. These wave fluctuations originate in the recirculation zone at the rear of the sphere, where two counter-rotating recirculation vortices alternately grow and separate. Strong flow instabilities occur in these regions, providing a continuous energy source for the wave fluctuations. These velocity disturbances propagate alternately along both sides of the sphere, forming a regular feature of the velocity disturbances. With respect to the spatial distribution, the wave packets of the velocity disturbances $u'$ and $v'$ are located between the two vortex legs, exhibiting an alternating pattern of peaks and valleys similar to that of hairpin vortices. These wave packets are arranged along the streamwise direction, forming a continuous multiple "S" waveform. At the same time, since the two vortex legs are vortices with opposite directions, the velocity disturbance $w'$ shows two consecutive multiple "S" waveforms with opposite phases along the streamwise direction, and its amplitude is relatively small. The hairpin vortex structure tends to dissipate in the downstream of the wake flow, but the wave packet structure generated by the velocity disturbances remains intact.

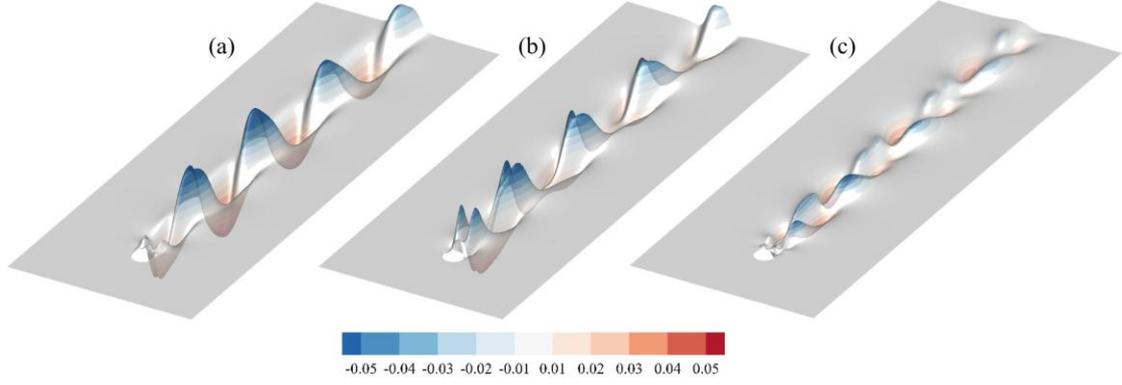

**FIG. 8.** Solitary wave structure identified by non-dimensional velocity disturbances at Re=280 (on y/d=-0.787 plane): (a) non-dimensional velocity disturbance, $u'/U_\infty$, (b) non-dimensional velocity disturbance, $v'/U_\infty$, (c) non-dimensional velocity disturbance, $w'/U_\infty$.

The velocity fluctuations in the three directions corresponding to the Reynolds numbers 280, 300 and 350 are presented in Fig. 9. And the monitoring points are represented as red dots in Figs. 7, 10 and 12, respectively. This location is one of the most unstable regions of the flow in and around the hairpin vortex[5]. At Re=280, the velocity fluctuations in three directions show synchronized cyclical variations, where the flow is almost two-dimensional and can be expressed by waves of u and v, exhibiting typical T-S wave characteristics. Respectively, the amplitudes of u, v and w velocity fluctuations are approximately 6.3%, 3.0% and 0.06% of the incoming velocity. At this time, the amplitude of u and v velocity fluctuations exceeds 2% of the incoming velocity, and the wake begins to enter the nonlinear collapse phase[1]. The three-dimensional characteristics of the velocity fluctuations under the current Reynolds number condition are significantly enhanced (amplitude increased by approximately 20 times) compared to those at Re=270. The 3D wave packet moves downstream at about 78% of the incoming velocity, which is similar to the SCS velocity measured by Jiang et al.[19] in the boundary layer flow. Jiang et al.[19] show that the moving speed of the SCS is about 60% - 80% of the incoming velocity. It can be seen that in the wake of a sphere, after the formation of the three-dimensional "kink" structures, it is hairpin vortices rather than Λ vortices that are commonly formed as the Reynolds number rises, and this is different from the transitional flow in the boundary layer.

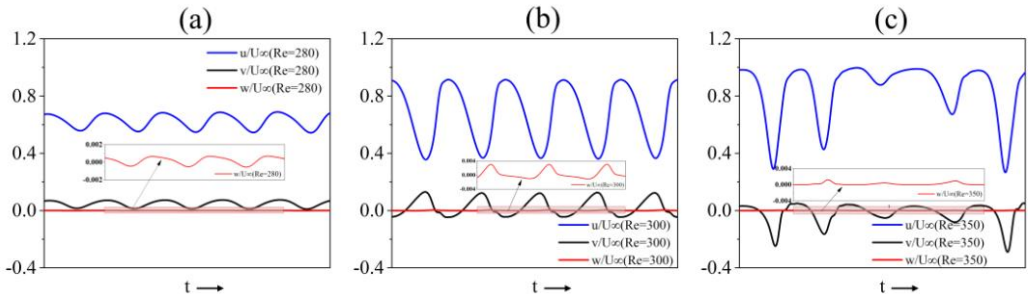

**FIG. 9.** Fluctuations of velocity components: (a) Re=280, T-S wave behavior, (b) Re=300, vortex shedding behavior, (c) Re=350, onset of velocity spike formation.

As it is illustrated in Fig. 10, at Re=300, the hairpin vortices show a mature periodic shedding to form a complete structure. The inclination angle of the hairpin vortices in the upper region is more significant than that in the lower region, which may be due to the difference in the dynamics of the fluid separation and reattachment of the sphere surface. A morphological feature that reveals significant differences in velocity and pressure gradients is that the upper hairpin vortices have smaller vortex heads and longer vortex legs. In Fig. 10, the red circles mark the vortex dislocation that occurs in the lower side hairpin vortex head. This result matches that of Tiwari et al.[32] under the same Reynolds number condition.

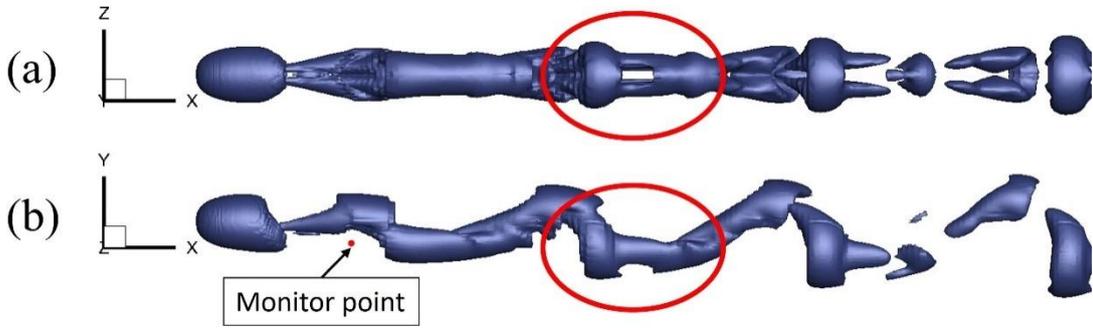

**FIG. 10.** Vortex structures identified by $\widetilde{\Omega}_R$ at Re=300, t=309s: (a) top view, (b) side view.

The 3D wave packets of the velocity disturbances at Re=300 are shown in Fig. 11. These wave packets appear sharper closer to the sphere compared to those at Re=280, while they gradually smooth out downstream. This shift in morphology suggests that the velocity fluctuations are undergoing a transition from T-S waves to spikes, corresponding to the characteristics of the velocity fluctuations recorded at the monitoring points in Figs. 9 (a) and 9 (b), which show a similar transition trend. The change in wave packet morphology and the change in fluctuation form at the monitoring points both suggest that the wake flow is undergoing a transition phase from laminar to turbulent. In Fig. 10, the position of the monitoring point in Fig. 9 (b) is marked with a red dot with respect to the sphere. At this monitoring point, the velocity component u shows regular periodic fluctuations in the time dimension, with a morphology between T-S waves and spikes. The streamwise velocity amplitude is kept at a level of about 40% of the incoming velocity at this position. The v and w components have a more regular morphology closer to that of T-S waves. Specifically, the amplitude of the velocity component v is approximately 7% of the incoming velocity, while the amplitude of the velocity component w is relatively smaller, around 0.1% of the incoming velocity.

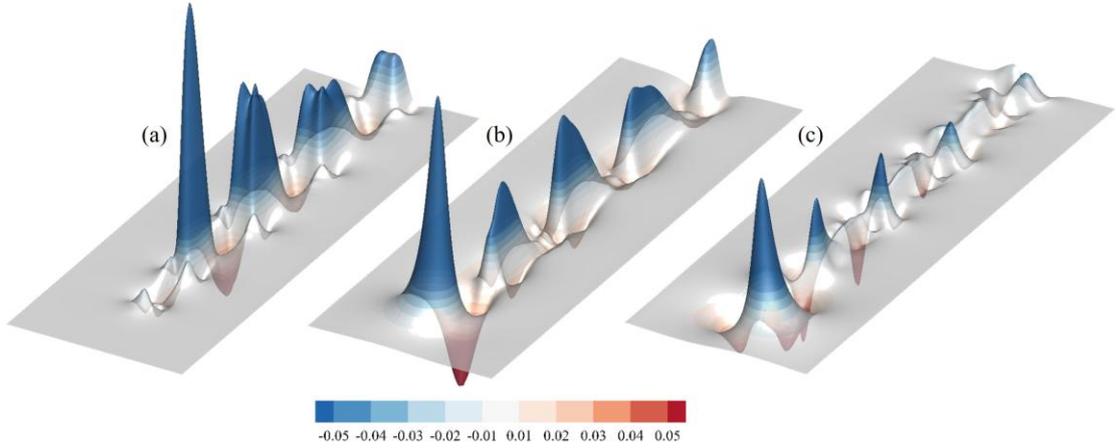

**FIG. 11.** Solitary wave structure identified by non-dimensional velocity disturbances at Re=300 (on y/d= -0.787 plane): (a) non-dimensional velocity disturbance, $u'/U_\infty$, (b) non-dimensional velocity disturbance, $v'/U_\infty$, (c) non-dimensional velocity disturbance, $w'/U_\infty$.

The structure of vortices in the wake identified by the $\widetilde{\Omega}_R$ method at Re=350 is shown in Fig. 12. It can be seen that these hairpin vortices maintain a high level of structural consistency with those at Re = 300, including similar inclination angle and vortex dislocation phenomena. However, the velocity fluctuation undergoes a significant shift. It can be seen from Figs. 11 and 13 that the wave packets at Re = 350 are more distinct compared to those at Re=300, especially within the recirculation zone in which the amplitude of the first wave packet grows by about 50%. Observed in the time sequence, the pattern of velocity fluctuations of u, v and w formally changes from a transitional pattern between T-S wave and spike at Re=300 to a spiky morphology at Re=350as shown in Figs. 9(b) and 9 (c). The negative peak of the u component can reach more than 60% of the incoming flow velocity. This is immediately followed by the positive peak of the v component (accompanied by an increasing velocity scalar), which can go up to more than 20% of the incoming velocity. The w component shows a positive peak that is just over 0.1% of the incoming velocity.

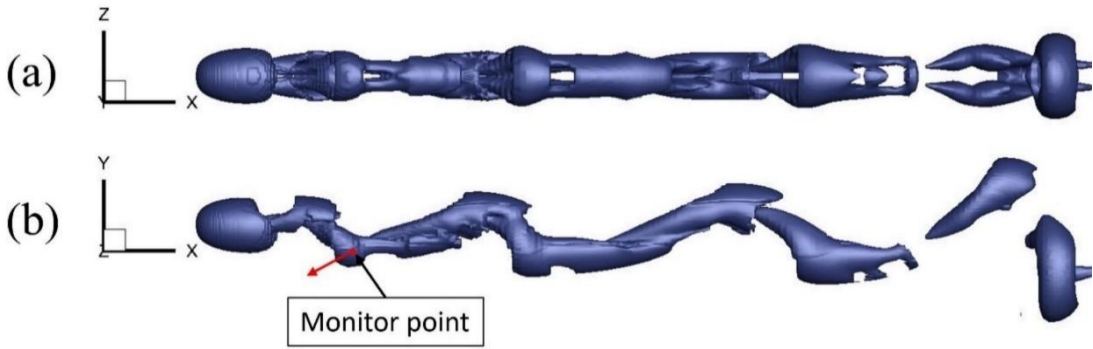

**FIG. 12.** Vortex structures identified by $\widetilde{\Omega}_R$ at Re=350:(a) top view, (b) side view.

The location of the negative velocity spike along the flow direction is the monitoring point which is marked as a red dot in Fig. 12. A velocity discontinuity takes place when the energy transfer between fluid layers ceases. Because of the viscosity of the fluid, the fluid element that has velocity discontinuity can't come to an immediate stop but rather decreases its velocity significantly, which leads to a negative spike. At this point, the flow is compressed along the flow direction, causing an almost simultaneous positive spike in the velocity v and triggering an ejection motion at the red point in Fig. 12 (b). The motion is in the direction shown by the red arrow, and this marks the start of turbulence generation. At the same time, as a result of the continuity equation and the conservation of mass, the velocity w exhibits a corresponding positive spike. From the perspective of the time sequence, the negative peak of the u component forms first. This is followed by the positive peak of the v component, and finally the positive peak of the w component. Consequently, the turbulence that originates from the spikes of u and v velocity components is generated through the ejection motion. The negative spikes of u play a key role and are the main driving force in sustaining velocity fluctuations. It is the component of velocity u that transfers the energy from the main flow to the fluctuations by its discontinuity (Dou[19]; Niu et al.[5]). The occurrence of these spikes denotes the onset of the turbulent transition in the wake. At Re=350, the formation of spikes in the wake gives rise to the sharpening of the 3D wave packets, which is a feature of the SCS. Such sharpening, which represents the singularity of the Navier-Stokes equation, indicates the beginning of the laminar-to-turbulent transition, and throughout this transition, the flow dynamics become more complicated.

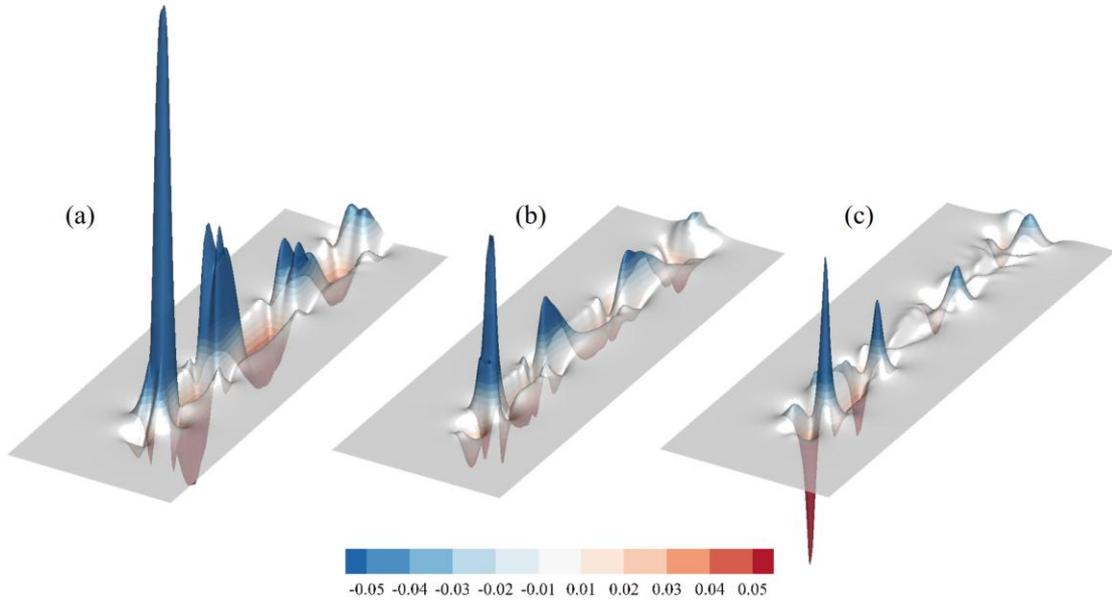

**FIG. 13.** Solitary wave structure identified by non-dimensional velocity disturbances at Re=350 (on the y/d= -0.787 plane): (a) non-dimensional velocity disturbance, $u'/U_\infty$, (b) non-dimensional velocity disturbance, $v'/U_\infty$, (c) non-dimensional velocity disturbance, $w'/U_\infty$.

Fig. 14 shows the relative positions of the hairpin vortex to the wave packet of the velocity disturbance $u'/U_\infty$ for different Reynolds numbers. When the Reynolds number is less than 270, the velocity disturbance $u'/U_\infty$ within the wake is weak and almost negligible. However, once the Reynolds number reaches 270, the amplitude of the velocity disturbance $u'/U_\infty$ is still small, but significant velocity fluctuations begin to appear with the appearance of the kink structure in the wake. When the Reynolds number goes up from 270 to 280, the velocity disturbance $u'/U_\infty$ in the wake begins to form a distinct regular wave packet. The wave packet of the negative velocity disturbance $u'/U_\infty$ is positioned in the middle of the two vortex legs. Furthermore, as the Reynolds number goes up from 280 to 300, the position of the wave packet of the velocity disturbance $u'/U_\infty$ changes significantly. In this evolution, the negative wave packet moves from the middle of the two hairpin vortex legs towards the central area of the hairpin vortex head, while its positive wave packet always occurs at the vertically symmetric position of the negative wave packet. With the continuous increase of the Reynolds number, the position of the negative wave packet in the central area of the hairpin vortex head remains unchanged. These features of the velocity disturbance, in particular the appearance of spikes and the stable relative position of the negative wave packet, together indicate the formation of the SCS structure.

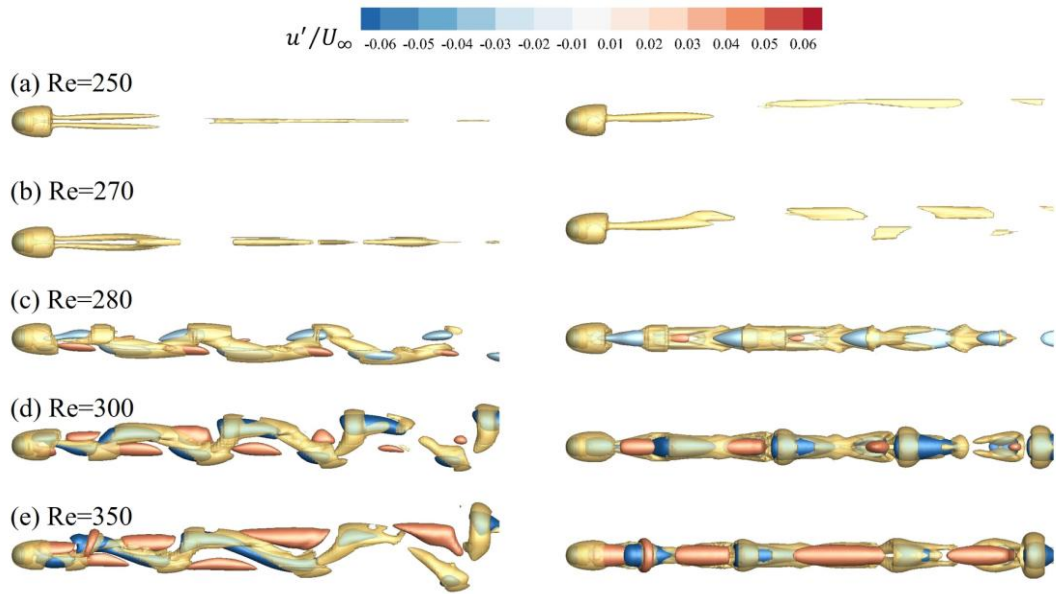

**FIG. 14.** Relative positions between the hairpin vortex and the wave packet of the velocity disturbance $u'/U_\infty$, vortex structures visualized by iso-surfaces of $\widetilde{\Omega}_R = 0.23$ in yellow, wave packets identified by iso-surfaces of $u'/U_\infty$ (negative in blue and positive in orange): (a) Re=250, (b) Re=270, (c) Re=280, (d) Re=300, (e) Re=350.

Kachanov et al.[43] showed that in the early stages of SCS formation, the structure develops mainly along the two flow directions (usually flow and spreading), with relatively little change in the third direction, which is normal to the flow direction. When they are in the two-dimensional phase, SCSs take on a spiky morphology. Additionally, these spikes rapidly recede from the wall as they form and propagate within the boundary layer with a relatively conserved morphology. This shows the conservatism of coherent structures, i.e., their morphology and amplitude remain almost constant during propagation in the boundary layer. Lee[2] has shown that within the SCS, the negative spikes in the flow direction are synchronized with the positive spikes in the propagation direction (y-axis). The three-dimensional features of the SCS within the wake of a sphere also experience a transition from a T-S wave morphology to a spiky morphology. As the Reynolds number rises from 280 to 350, the velocity fluctuation changes from a T-S wave morphology to a spiky morphology. At Re=350, the velocity spikes in the three flow directions occur almost simultaneously. First there is a negative spike in the u component, followed by a positive spike in the v component, and finally a positive spike in the w component. The spike in the third direction was not discussed in the studies of Kachanov et al.[43] and Lee[2].

## 3.3 SCS evolution

The instability in the recirculation zone behind the sphere triggers the velocity fluctuation at Re=270, and this velocity fluctuation is also one of the key factors for the kink formation. In the Reynolds number range from 280 to 350, the velocity fluctuation is significantly enhanced, and its morphology undergoes a transition from a T-S wave to a sharper and spiky wave. This transition has a substantial impact on the evolution of the vortex structures. The enhanced velocity fluctuation produces higher shear around the head of the hairpin vortex, which allows the vortex head to last longer than the vortex legs. In addition, the wave packet of the velocity disturbance shows significant regularity and synchronization with the hairpin vortex structure. It is revealed that there is an orderly interaction between the velocity disturbance and the vortex structure at the onset of the turbulent transition, and their relative positions follow a predictable pattern. The solitary wave properties of the velocity disturbances in the time sequence at higher Reynolds numbers are further analyzed below.

Two monitoring lines are defined. Line 1 is positioned at those points where the values of (x/d, y/d, z/d) span from (4, 0, -4) to (4, 0, 4). In contrast, Line 2 is located at the points where the (x/d, y/d, z/d) values range from (0, 4, 0) to (0, -4, 0). The configuration makes it possible to monitor and analyze how the flow goes in both directions, and the positions of the two lines relative to the vortex structure are presented in Fig. 15.

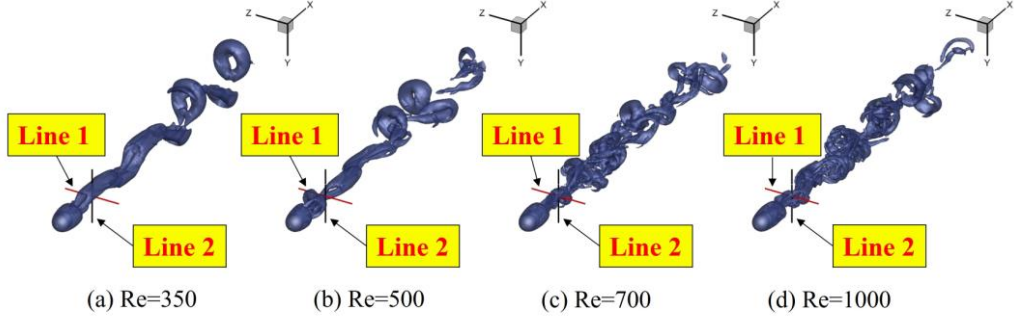

**FIG. 15.** Position of monitor lines: (a) Re=350, (b) Re=500, (c) Re=700, (d) Re=1000.

The time evolution of the velocity disturbance along the two monitoring lines as well as the corresponding Reynolds numbers are presented in Fig. 16. At the Reynolds number of 350, the monitoring Line 1 is at right angles to the symmetry plane of the hairpin vortex, while the monitoring Line 2 is on this symmetry plane. The time trajectory of the velocity disturbances $u'/U_\infty$ along Line 1 clearly shows the morphology of the solitary wave. These wave packets of the velocity disturbance $u'/U_\infty$ exhibit periodic fluctuations up and down along the time axis, which directly reflects the property of the solitary wave to maintain its morphology and amplitude unchanged during propagation. Monitoring Line 2 passes through the hairpin vortex legs. As shown in Fig. 16 (e), the velocity disturbances $u'/U_\infty$ at the two symmetric positions of y/d=-0.5 and y/d=0.5 show periodic fluctuations with alternating peaks and valleys, as well as two sets of periodic up-and-down motions along the time axis. This pattern is consistent with the spatial pattern of the velocity disturbances $u'/U_\infty$ at this Reynolds number, and reflects the fact that the wave packet of the velocity disturbance $u'/U_\infty$ in the wake has the properties of a solitary wave in both time and space.

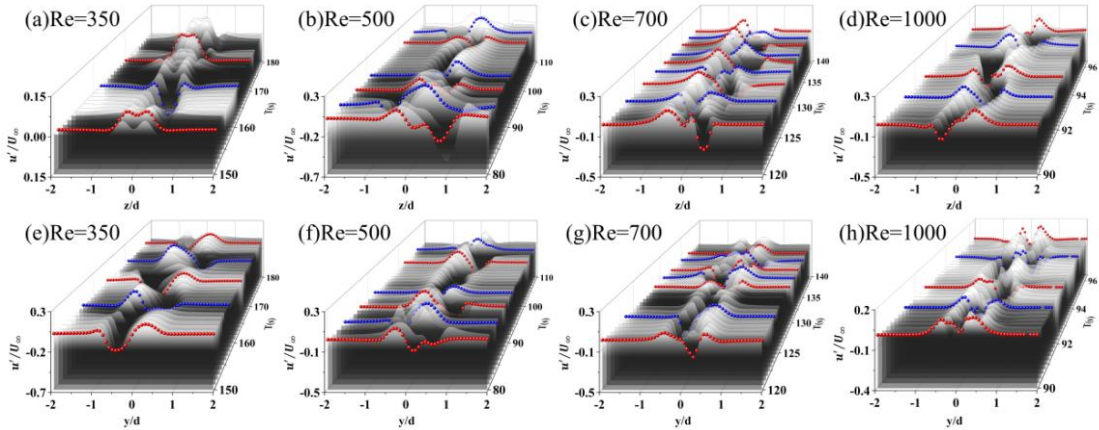

**FIG. 16.** Solitary wave structure in the wake, $u'/U_\infty$ distribution along Line 1: (a) Re=350, (c) Re=500, (e) Re=700, (g) Re=1000, $u'/U_\infty$ distribution along Line 2: (b) Re=350, (d) Re=500, (f) Re=700, (h) Re=1000.

As the Reynolds number rises to 500, the hairpin vortex sheds and rotates around the flow axis, breaking its original planar symmetry, and small vortex structures begin to form around the primary hairpin vortex. These changes affect the relative positions of the monitoring Line 1 and Line 2 with respect to the hairpin vortices, causing them to be no longer fixed, as shown in Fig. 16. Correspondingly, the complexity of the velocity disturbance $u'/U_\infty$ in the wake increases. Once the wave packets of the velocity disturbances exit the recirculation zone, they are affected by not just the "ejection" and "sweep" motions of the primary hairpin vortices, but also by the corresponding motions of the small vortices encircling the primary hairpin vortices. This complex interaction complicates the velocity disturbance along the monitoring lines Line 1 and Line 2. For a given level of the Reynolds number within the range from 350 to 1000, the frequency at which the primary hairpin vortex sheds from the sphere remains relatively constant[40, 44]. In addition, the wave packet of the negative velocity disturbance $u'/U_\infty$ constantly emerges at the center of the head of the primary hairpin vortex, such as the blue wave packets in Fig. 17. This indicates that the negative wave packet of the velocity disturbance $u'/U_\infty$ also leaves the recirculation zone at the same frequency as the shedding frequency of the hairpin vortex, and is able to maintain its shape and amplitude unchanged in the near wake region. This phenomenon further confirms that from the Reynolds number 350 to 1000, the wave packet of the negative velocity disturbance $u'/U_\infty$ still maintains regular at a given Reynolds number level. Its shape and amplitude remain almost constant in the near wake region, showing solitary wave characteristics.

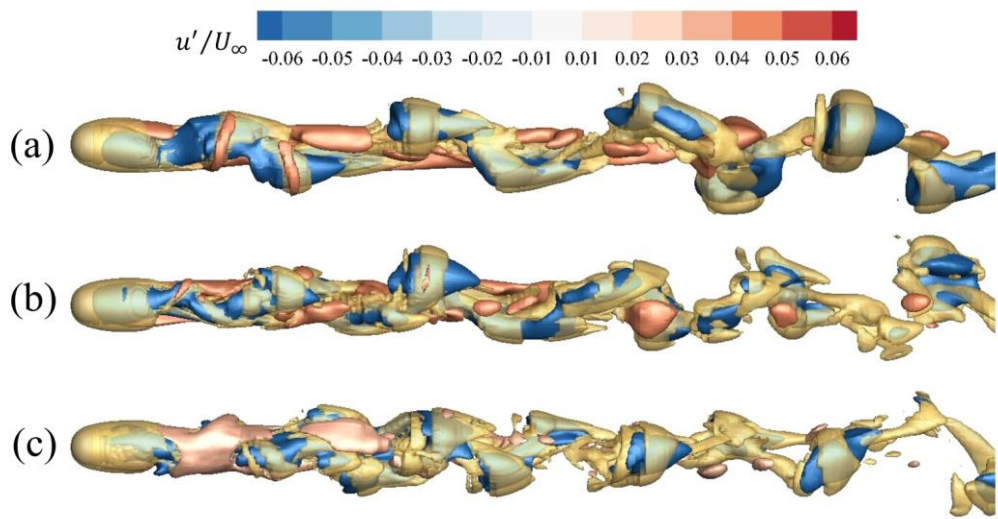

**FIG. 17.** Relative positions of the vortex structures to the wave packet of the velocity disturbance $u'/U_\infty$, vortex structures visualized by iso-surfaces of $\widetilde{\Omega}_R = 0.23$ in yellow, wave packets identified by iso-surfaces of $u'/U_\infty$ (negative in blue and positive in orange): (a) Re=500, (b) Re=700, (c) Re=1000.

For different Reynolds numbers (Re=350, 500, 700, and 1000), Fig. 18 shows the velocity fluctuations in the wake at a monitoring point located in the x/d=3.937 plane. This monitoring point is located directly on the side of the wake, at the lower edge of the head of a primary hairpin vortex, which is one of the most unstable regions of the flow around the hairpin vortex[5]. The "ejection" motion formed by the velocity fluctuations can be identified in Fig.18. In the initial phase of the turbulent transition of the wake (Re=350), from the time sequence analysis, the velocity component u (direction of flow) first shows a negative spike, followed by the velocity component v (vertical direction of flow) and finally the velocity component w (direction of spreading), with a sequence as shown by markers 1, 2 and 3 in Fig. 18(a). The magnitude of the spike shows that the negative spike of u is the most significant, followed by v, while the magnitude of w is relatively small. This difference in time sequence and amplitude suggests that the negative spike of u component is key and dominant in the turbulent transition. It is the main driving force for the onset of turbulence generation.

In the study that focused on the large-scale turbulent structure generated by a broken solitary wave on a 1/50 plane slope, Ting[45] found significant correlations among the velocity disturbance components $u'$, $v'$, and $w'$. Moreover, these obtained correlations reveal the influence of coherent structures on turbulence characteristics. Feng et al.[16] made a discovery in their study on spike-like solitary waves which are driven by a travelling wave in an incompressible boundary layer. They found that as the amplitude of the driving wave gradually increases, the flow structure experiences a transition from being regular to becoming irregular. Moreover, it was observed that the velocity u component holds the majority of the fluctuating energy within the solitary wave, and the $v'$ and $w'$ components are related to the three-dimensional properties of the flow. The key role of the negative spikes of u in the turbulence transition is supported by these results. The uncertainties of the velocity components v and w change significantly with increasing Reynolds number. When there is a negative spike in the streamwise velocity u, the spikes in the velocity components v and w tend to be either positive or negative, as depicted in Fig. 18 (b), 18 (c), and 18 (d). Additionally, the dynamic behaviors of the velocity components u, v, and w must conform to the fundamental principles of fluid mechanics, such as the continuity equation and the law of conservation of mechanical energy, so as to guarantee physical consistency.

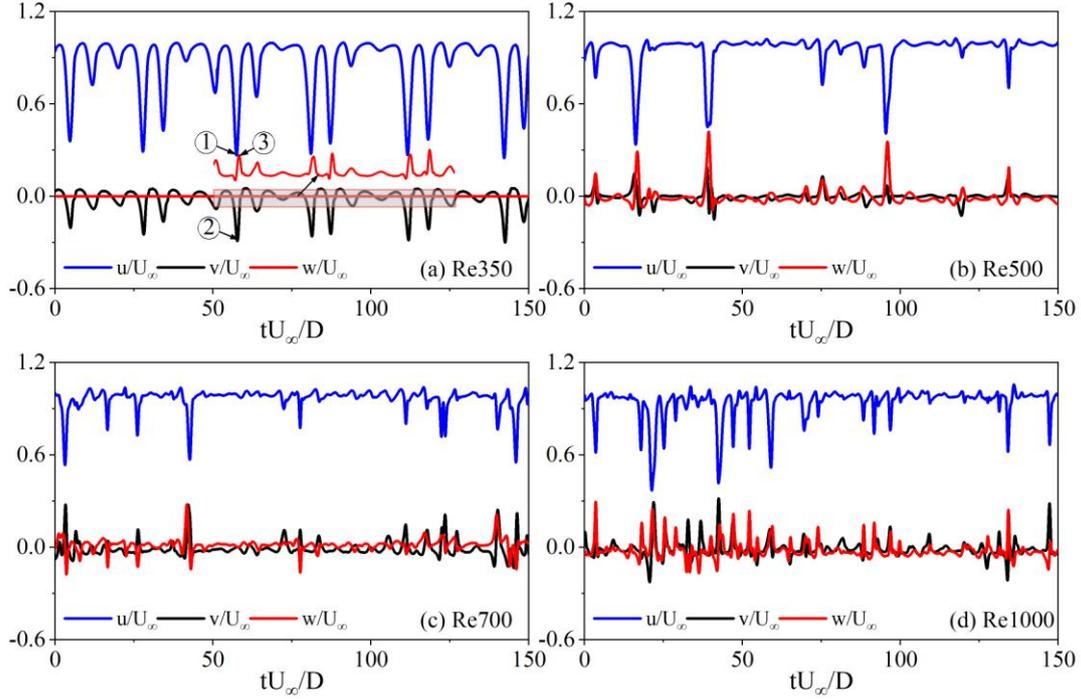

**FIG. 18.** Velocity fluctuations with time at fixed monitoring points on the x/d=3.937 plane where the "ejection" motion can be identified. (a) Re=350, (b) Re=500, (c) Re=700, (d) Re=1000.

Table II presents a summary of the evolution of the SCS. Specifically, at Re=250, the vortex structure is shown as two parallel flow vortices accompanied by weak pulsations, and the three-dimensional wave packet has not yet formed. As the Reynolds number increases to 270, the kink structure appears in the wake, accompanied by a weak three-dimensional wave packet with a velocity of about 48% of the incoming velocity. The Reynolds number goes up further to 280. At this time, the hairpin vortices begin to shed, but are not yet fully developed, and the velocity fluctuations in the wake are manifested as T-S waves, with the three-dimensional wave packet exhibiting a regular fluctuation pattern of alternating single peak and single valley, at a velocity of about 78% of the incoming velocity. As the Reynolds number continues to increase, the hairpin vortex begins to shed regularly when the Reynolds number is between 300 and 350, and the SCS maintains an alternating pattern of single peak and single valley in space (as shown in Fig. 19 (a)), while in time the velocity fluctuation transforms from a T-S wave to a spiky morphology, marking the formal formation of the SCS. In the range of Re between 350 and 1000, the hairpin vortex oscillates and deforms, forming secondary vortex structures around it, making the SCS more complex. The fluctuation pattern also changes from a regular single peak alternating with a single valley to a complex pattern with multiple peaks and valleys (shown in Fig. 19 (b)). This complex pattern of fluctuations is not only more morphologically complex, but also shows an overall downstream movement during propagation, with propagation speeds of about 80% to 90% of the incoming velocity.

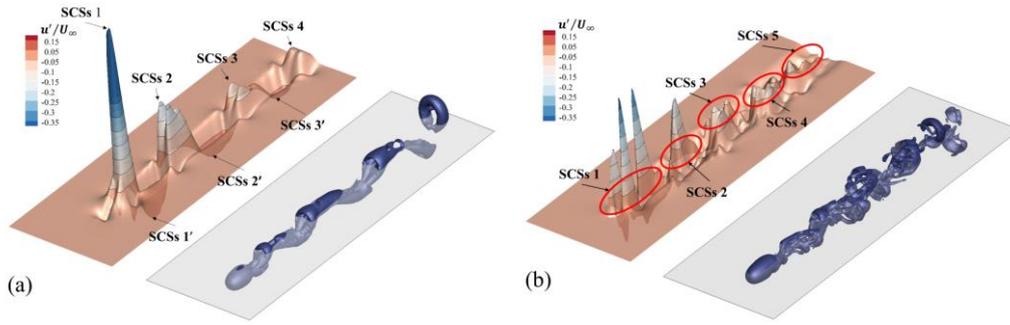

**FIG. 19.** Relative position between the SCS (identified by $u'/U_\infty$ on the y/d=-0.787 plane) and vortex structures (identified by $\widetilde{\Omega_R} = 0.23$): (a) Re=350, (b) Re=1000.

**Table II** Evolution and characterization of the SCS with increasing Reynolds number.

| Re | Vortex structure | Velocity fluctuation | 3D wave packet | SCS | SCS speed relative to the incoming velocity |
|---|---|---|---|---|---|
| 250 | Parallel streamwise vortices | Faint TS wave | / | / | / |
| 270 | "kink" appears at around x/D=6 | Faint TS wave | Exist | Early pattern | 48% |
| 280 | Hairpin vortices | TS wave | Exist | Early pattern | 78% |
| 300 | Hairpin vortices, vortex dislocation | Transitional form between TS wave and spike | Exist | Early pattern | 81% |
| 350 | Hairpin vortices, vortex dislocation | Spike | Exist | Exist | 81% |
| 500 | Asymmetrical hairpin vortices, secondary vortices | Spike | Exist | Exist | 79% |
| 700 | Asymmetrical hairpin vortices, secondary vortices | Spike | Exist | Exist | 90% |
| 1000 | Asymmetrical hairpin vortices, secondary | Spike | Exist | Exist | 87% |

vortices

## 3.4 SCS/High-shear layer/Hairpin vortex

The current analyses reveal that the SCS has a key part to play in the turbulent transition of the wake flow. Moreover, the formation and evolution of its 3D wave structure are closely associated with the vortex structures. The evolution of the SCS experiences three stages as depicted in Fig. 20.

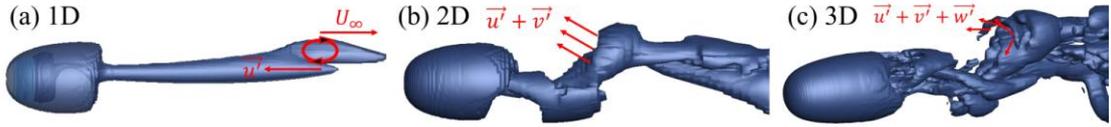

**FIG. 20.** Three phases of the SCS effect on vortices by velocity fluctuations: (a) one-dimensional effect, Re=270, (b) two-dimensional effect, Re=350, (3) three-dimensional effect, Re=1000.

During the first stage when the Reynolds number is low, its three-dimensional features (see 3.1, amplitude of the velocity components at Re=270) are not prominent, and this impact is primarily exhibited as a one-dimensional effect in the streamwise direction. The instability of the recirculation zone behind the sphere gives rise to this fluctuation. As a result, there are fluctuations in the velocity difference between the high-speed fluid of the main flow and the low-speed fluid of the wake flow along the flow direction. As shown in Fig. 20 (a), this fluctuation shear at locations with large amplitudes, which promotes the formation of kink structures. The formation of kink structures is the result of velocity fluctuations caused by the instability of the recirculation zone.

During the second stage, with the Reynolds number rising to 350, the instability of the recirculation zone is intensified (see 3.2, amplitude of the velocity components at Re=350). The SCS transitions from one-dimensional effect to two-dimensional effect, although the fluctuation of the third dimension (z-direction) is still small, less than 0.2% of the incoming velocity. At this stage, the velocity fluctuation in the wake undergoes a change from a T-S wave to a spiky morphology, which marks the formal formation of the SCS.As depicted in Fig. 12 (b), in the XY plane, the velocity disturbances in the X and Y directions play a dominant role in the ejection motion. They take the low-speed wake flow fluid into the high-speed main flow zone, make fluid mix better and form a big high-shear layer at the main flow and wake junction, as shown in Fig. 20(b). At the same time, sweep phenomena occur on both sides of the vortex legs, and these ejection and sweep motions are the energy sources for the generation and development of turbulence. These velocity fluctuations interact with the ejection and sweep motions that are brought about by the primary hairpin vortex. Collectively, they influence the fluid and give rise to a robust high-shear layer, which provides favorable circumstances for the creation of secondary vortex structures in the upstream region of the primary hairpin vortex.

During the third stage, with the Reynolds number continuing to increase, the three-dimensional effect of the SCS starts to become evident. Meanwhile, the velocity disturbances in the three directions (see 3.3, amplitude of the velocity components from Re = 500 to Re = 1000) collaborate with each other to give rise to the instability of the high-shear layer, and this, in turn, triggers the instability of the fluid. It is shown in Fig. 20 (c) that the rolling up of the high-shear layer gives rise to a small secondary vortex structure upstream of the primary hairpin vortex. The occurrence of secondary vortex structures agrees with the findings of Zhou et al.[46] in channel flow. Therefore, the turbulent transition of the wake flow is the consequence of the temporal and spatial evolution of the velocity fluctuations generated within the recirculation zone.

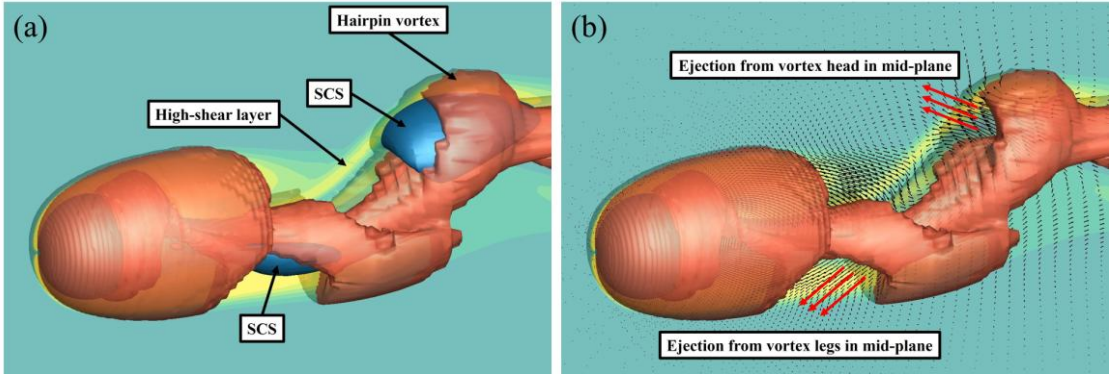

**FIG. 21.** Relative position: vortex structures (identified by $\widetilde{\Omega_R}$ method), SCS (identified by $u'/U_\infty = -0.2$) and high-shear layer (contours of vorticity magnitude on XY plane, high-shear in yellow), Re=350, t=135s.

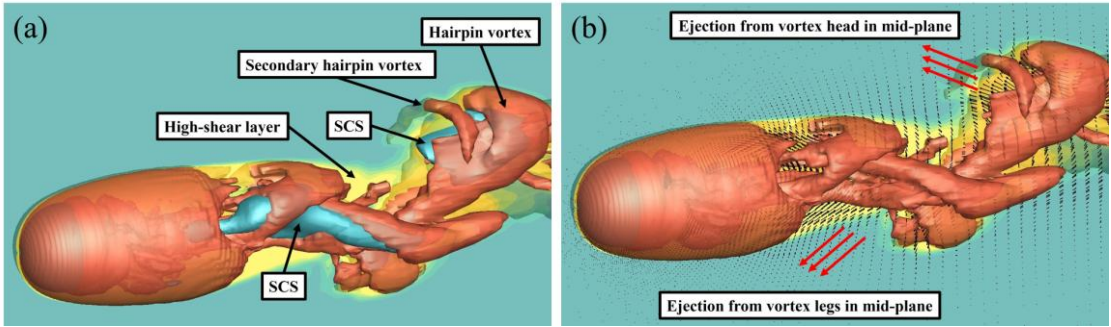

**FIG. 22.** Relative position: vortex structures (identified by $\widetilde{\Omega_R}$ method), SCS (identified by $u'/U_\infty = -0.2$) and high-shear layer (contours of vorticity magnitude on XY plane, high-shear in yellow), Re=1000, t=98s.

The hairpin vortex is located at the periphery of the SCS in the wake of a sphere, and the high-shear layer wraps further around the outside of the hairpin vortex, as shown in Figs. 21 and 22. This structural arrangement is similar to the phenomena observed in boundary layer flow by Lee[2]. In these studies, it was found that the high shear layer and vortex structures usually form and then evolve along the borders of the SCS. This suggests that the hydrodynamic behavior follows a similar pattern in both boundary layer flow and wake flow, where the SCS has an important influence on the flow properties.

## 4. Conclusions

(1) **Appearance of the initial wave packets**. In the initial phase of the turbulent transition, two alternating unique three-dimensional wave packets are found in the wake of a sphere. This flow structure marks the initial appearance of the SCS. This structure is formed before the vortex shedding in the T-S wave stage of the wake flow. In the initial stage (Re=270), a shear layer is formed at the junction due to the velocity difference. The instability within the recirculation zone following the sphere results in periodic velocity fluctuations and contributes to the formation of the three-dimensional wave packets. As for the wave packet between the two streamwise vortices, it accelerates the roll-up of the shear layer, which ultimately leads to the formation of a kink structure. The appearance of the kink denotes the beginning of the formation of the wave packet featuring velocity pulsation, which is the incipient stage of the SCS. The amplitude of the three-dimensional wave packet is rather low in this stage.

(2) **Soliton-like coherent structure.** The SCS in the wake of a sphere is in the form of a three-dimensional wave packet. This wave packet brings about a negative spike in the streamwise velocity and is located right at the head center of the hairpin vortex. The three-dimensional nature of this structure is mainly reflected in the distribution of velocities. As for the negative spike of the streamwise velocity u, it is caused by a velocity discontinuity, at which point the energy transfer among the fluid layers ceases. Because of the fluid's viscosity, the fluid element in which the velocity discontinuity occurs cannot be stopped immediately, but is instead transitioned by a sharp decrease in velocity, resulting in the negative spike. The position of this discontinuity gives rise to a singularity in the Navier-Stokes equations. The superposition of the velocity vector u and the velocity vector v in the central plane of the hairpin vortex produces an ejection motion zone. This ejection motion produces a high shear stress in the Q2 quadrant of the two-dimensional velocity fluctuations[43]. Subsequently, a high-shear layer is formed at the junction of the high-speed fluid and the low-speed fluid, which is precisely the exterior of the hairpin vortex. As the high-shear layer increases, so does its instability, resulting in intense fluid instability. In addition, the high-shear layer rolls up to form a secondary vortex structure.

(3) **Relative position of the SCS and hairpin vortex**. In the SCS's initial stage, the wave packet of the negative streamwise velocity is located between the two streamwise vortices, while the positive wave packet is located at the position of the kink structure. When the velocity spikes are generated in the wake, the position of the SCS moves with the Re increase, moving from the middle of the two hairpin vortex legs towards the central area of the hairpin vortex head. This state maintains until the Reynolds number reaches 1000. During this evolution, the high-shear layer always covers the surrounding area of the hairpin vortex.

(4) **Power for turbulent generation.** In the early stage of the turbulent transition, from the time sequence, the negative spike of the u component is formed first, followed by the positive spike of the v component, and finally the positive spike of the w component. The magnitude of the spike indicates that the negative spike of the u component is the most significant, followed by the v component, while the magnitude of the w component is relatively small. This difference in time sequence and amplitude suggests that the negative spike of the u component is key and dominant in the turbulent transition. It is the main driving force for the onset of turbulence generation.

(5) **The fluctuation pattern of the SCS varies with Reynolds number.** When the Reynolds number is between 280 and 300, the velocity fluctuation takes the form of a T-S wave. As soon as the Reynolds number exceeds 300, the velocity fluctuation in the wake begins to show significant spikes until the wake changes completely to a turbulent state. At the stage where the hairpin vortex has not yet formed a secondary vortex structure (280<Re<350), the SCS in the wake shows a regular pattern, i.e. a single peak alternating with a single valley. When the Reynolds number further goes up, the primary hairpin vortex begins oscillating and distorting, while the secondary vortex structures begin to form around it. This change leads to a more complex structure of the SCS and the fluctuation pattern changes from the original regular single peak and single valley alternating to a complex multi-peak and multi-valley fluctuation pattern. This multi-peak and multi-valley pattern is morphologically more complex and moves downstream with a propagation speed of about 80%-90 % of the incoming velocity.

(6) **Effects of the SCS**. There are three stages in the evolution of the SCS. The first stage is the one-dimensional effect, which represents the formation of the kink structure and the appearance of 3D wave packets. The second stage is the two-dimensional effect, which indicates the appearance of a high shear layer. The third stage is the three-dimensional effect, reflecting the formation of secondary vortices.